\documentclass[a4paper,fleqn,usenatbib]{mnras}

\usepackage{newtxtext,newtxmath}

\usepackage[T1]{fontenc}
\usepackage{ae,aecompl}

\usepackage{graphicx}	
\usepackage{amsmath}	
\usepackage{mathptmx}
\usepackage{amssymb}	

\title[The halo assembly from APOGEE.]{The assembly history of the Galactic inner halo inferred from $\alpha$-element patterns }

\author[Fern\'andez-Alvar et al.]{
Emma Fern\'andez-Alvar,$^{1,2}$\thanks{E-mail: emmafalvar@gmail.com}
Patricia B. Tissera,$^{3,4}$
Leticia Carigi,$^{2}$
\newauthor
William J. Schuster,$^{5}$,
Timothy C. Beers$^{6}$
and Vasily A. Belokurov$^{7,8}$
\\
$^{1}$Universit\'e Cote d'Azur, Observatoire de la Cote d'Azur, CNRS, 06300, Laboratoire Lagrange\\
$^{2}$Instituto de Astronom\'ia, Universidad Nacional Aut\'onoma de M\'exico, AP 70264, 04510, Ciudad de M\'exico, M\'exico\\
$^{3}$Departamento de Ciencias F\'isicas, Universidad Andr\'es Bello, Av. Rep\'ublica 220, Santiago, Chile\\
$^{4}$Millenium Institute of Astrophysics, 700 Fern\'andez Concha, Santiago, Chile\\
$^{5}$Observatorio Astron\'omico Nacional, Universidad Nacional Aut\'onoma de M\'exico, Apartado Postal 877, CP 22800, Ensenada, BC, M\'exico \\
$^{6}$Department of Physics and JINA Center for the 
Evolution of the Element (JINA-CEE),
University of Notre Dame,\\
 Notre Dame, IN 46556 USA\\
$^{7}$Institute of Astronomy, Madingley Rd, Cambridge, CB3 0HA \\
$^{8}$Center for Computational Astrophysics, Flatiron Institute, 162 5th Avenue, New York, NY 10010, USA \\
}

\date{Accepted XXX. Received YYY; in original form ZZZ}

\pubyear{2017}
\begin{document}
\label{firstpage}
\pagerange{\pageref{firstpage}--\pageref{lastpage}}
\maketitle

\begin{abstract}

We explore the origin of the observed decline in [O/Fe] (and [Mg/Fe]) with
Galactocentric distance for high-metallicity stars ([Fe/H] $> -1.1$),
based on a sample of halo stars selected within the Apache Point Observatory
Galactic Evolution Experiment (APOGEE) fourteenth data release (DR14).
We also analyse the characteristics of the [$\alpha$/Fe] distributions in the
inner-halo regions inferred from two zoom-in Milky Way mass-sized
galaxies that are taken as case studies. One of them qualitatively reproduces
the observed trend to have higher fraction of $\alpha$-rich star for
decreasing galactocentric distance; the other exhibits the
opposite trend.  We find that stars with [Fe/H] $> -1.1$ located in the range
[15 - 30] kpc are consistent with formation in two starbursts, with maxima
separated by about $\sim1$ Gyr.  We explore the contributions of stellar populations with
different origin to the [$\alpha$/Fe] gradients detected in stars with
[Fe/H] $> -1.1$. Our analysis reveals that the simulated halo that best
matches the observed chemical trends is characterised by an accretion
history involving low to intermediate-mass satellite galaxies with a short
and intense burst of star formation, and contributions from a more massive
satellite with dynamical masses about
$\sim 10^{10}M_{\sun}$, distributing low [$\alpha$/Fe] stars at
intermediate radius.

\end{abstract}

\begin{keywords}
Galaxy: formation -- Galaxy: halo -- Galaxy: abundances -- methods: numerical -- techniques: spectroscopic
\end{keywords}



\section{Introduction}
Observations of the Milky Way (MW) and the haloes of nearby galaxies have
revealed spatial sub-structures \citep{majewski2003,belokurov2006,
morrison2009,bonaca2012, sesar2012, gilbert2013,ibata2014}, as well as
differences in the kinematical and chemical properties of the stars in
the inner region of the Galactic halo with respect to the outer region
\citep{carollo2007,carollo2010,beers2012,fernandezalvar2015}. These results
suggest that the stellar populations dominating the inner- and outer-halo
regions formed in systems with different star-formation histories. One explanation proposed is that the outer halo was
built mainly from mergers of satellite galaxies, whereas the inner
halo might have received contributions of stars formed in-situ in the Galaxy \citep[for example stars from a heated disc due to merger events, e.g. ][]{sheffield2012, nissen2010, schuster2012}



The formation of the stellar halo has been tackled using dynamical and
hydrodynamical simulations \citep[e.g.,][]{bullock2010,cooper2010,
tissera2013,pillepich2015}. All of these works predict the formation of
the stellar halo mainly from the contribution of satellites with a
variety of masses. However, hydrodynamical simulations also reported the
possible existence of an in-situ component \citep{zolotov2009,
tissera2012, cooper2015}. Disc-heated stars are a possible candidate to
explain the origin of the in-situ components \citep[as proposed from both numerical simulations as well as observational works, e.g.][]{tissera2014,
bonaca2017}, but they might have also formed from gas transported in by
gas-rich satellites \citep{zolotov2009, tissera2013}.
\citet{harmsen2017} found that these simulations were able to reproduce
global trends of the stellar haloes of nearby galaxies, although most of
them predicted more massive stellar haloes than inferred from
observations, and larger contributions of in-situ stars compared to the
observed estimates.

However, recently observations have shown evidence for the accretion of a
relatively massive satellite contributing stars to the inner halo
of the Milky Way. \citet{belokurov2018} observed that the orbits of stars with [Fe/H]
$> -1.7$ are strongly radial, in a sausage-like distribution, and, by comparing with N-body simulations,
deduced that they come from the merger of a satellite with virial mass M
$> 10^{10}$ M$_{\odot}$, which was called in following works the ``Gaia sausage". Using
data from the Gaia
second data release (DR2), \citet{deason2018} obtained that stars with
very radial orbits (like in the ``Gaia sausage") have apocentres which
pile up at $r \sim$ 20 kpc, where the spatial break in the halo was
measured in numerous previous papers. They conclude that these stars likely come
from a satellite with stellar mass at least M $> 10^{8}$ M$_{\odot}$, or
from a group of dwarf satellites accreted at similar times.
\citet{helmi2018} identified stars in retrograde orbits, and they
also constrained \citep[from the star-formation rates obtained from similar
stars in][] {fernandezalvar2018a} the stellar mass of their
progenitor satellite to be $\sim 6\times 10^{8}$ M$_{\odot}$ (which they
called ``Gaia-Enceladus"). 

From the
analysis of globular clusters in the Galactic halo, other evidence was found
for relatively massive mergers. \citet{myeong2018} confirmed
that the configuration of a group of globular clusters in action space
is compatible with their coming from the ``Gaia sausage" merger event.
\citet{kruijssen2018} used the age-metallicity relation in halo
clusters, compared with the E-MOSAICS hydrodynamical simulations, to
infer that the MW may have undergone mergers with three satellites with
stellar masses between $10^{8}$ and $10^{9}$ (the most massive of which
they refer to as ``Kraken"). Other evidences have been also provided by \citet{lancaster2018,haywood2018,fernandezalvar2018b,simion2018,iorio2018,wegg2018,mackereth2018}.

An important piece of information is provided by $\alpha$-element
abundances \citep[e.g.,][]{zolotov2009}. Using the Aquarius haloes,
\citet{tissera2013} showed that stars in the stellar haloes have
different $\alpha$-enrichments that could provide hints on the mass and
star-formation history (SFH) of the accreted satellites, as well as the
relative contributions of stars formed through different processes. In
their simulations, the authors identifed accreted stars (or debris)
formed in systems outside the virial radius of the main galaxy
progenitor, and later on, in-situ stars, formed within the virial
radius. Within the latter population, they also identified the so-called
endo-debris stellar populations, formed from gas carried in by accreted
satellites, and disc-heated stars, born in-situ in the disc and later,
dynamically heated into the halo. Their different origin predicts
different chemical and kinematical patterns in the inner regions of the
stellar haloes, in particular different $\alpha$-enhancement patterns.


Indeed, a promising path for investigation is consideration of the
[$\alpha$/Fe] abundance ratios that have been derived from observations
of large numbers of stars in the MW stellar halo
\citep{fernandezalvar2015, fernandezalvar2017}. The $\alpha$-elements
are ejected into the interstellar medium (ISM) on short timescales
(Myr), because these elements are produced mainly by massive stars that
explode as Type II supernovae (SNeII). In contrast, release of the bulk
of Fe occurs on longer timescales (Gyr), because iron is synthesised by
binary stars of low and intermediate mass that explode as Type Ia
supernovae (SNeIa) . The patterns in the relative abundances
[$\alpha$/Fe]\footnote{[$\alpha$/Fe] = $\log$($\alpha$/Fe) $-
\log$($\alpha$/Fe)$_{\odot}$} vs. [Fe/H]) are well-known to be affected
by the star-formation rate and the initial mass function (e.g., see Fig.
1 of \citet{fernandezalvar2018a}).


\citet{fernandezalvar2017} identified abundance gradients for
$\alpha$-elements and other chemical species relative to the iron
abundance with distance from the Galactic center, based on an analysis
of over 400 stars from the APOGEE twelfth data release (DR12). They
found that the nature of these gradients depended on the [Fe/H] range,
flattening as [Fe/H] decreases, suggesting a variation of the
chemical-enrichment history for stars as a function of distance. The
identification of Galactic halo simulations able to reproduce the
observed trends can help to unravel the assembly history of our Galaxy.

In this work, we examine the [$\alpha$/Fe] trends with Galactocentric
distance observed in the halo of the MW by the APOGEE fourteenth data
release (DR14), which has tripled the number of stars with this
information available. We compare them with the trends inferred from a
set of simulated haloes of MW mass-size galaxies in order to constrain
the assembly history that led to the stellar populations in the
inner-halo region with the observed chemical compositions.

We employ the set of Aquarius haloes analysed by \citet{tissera2012,
tissera2013,tissera2014} and \citet{carollo2018} that correspond to MW mass-size
galaxies to study the chemical enrichment of baryons during Galactic
assembly. These authors found that the stellar halo of the MW is characterised by
combination of stellar populations with different chemistry, kinematics,
ages, and binding energies depending on their origin. We acknowledge the
fact that the Aquarius haloes \citep{scan09} are known to over-produce
stars at high redshift. However, they comprise a diversity of assembly
histories for galaxies of similar virial masses that allow us to
explore which assembly history provides the best representation of the
observations. Our approach is to use them as case studies to identify 
the main processes that could explain the origin of the
$\alpha$-abundance patterns detected in observations of the MW's halo. 

This paper is organised as follows. In section 2 we describe the main
characteristics of the observational and simulated data used in this
work. In section 3 we infer the chemical trends from both sets, and
analyse their similarities and differences. Finally, in section 4 we
summarise our main results and conclusions.

\section{Observational and simulated  data}


\subsection{Observations}
We make use of the APOGEE fourteenth data release, DR14
\citep{blanton2017}. This stellar spectra database includes all the
APOGEE observations gathered from May 2011 to July 2014
\citep{eisenstein2011}, and the observations from the first two
years of its extension, APOGEE-2. Using a multi-object infrared
spectrograph \citep{wilson2010} coupled to the 2.5-meter telescope at
the Apache Point Observatory \citep{Gunn2006}, high-resolution
($R\sim22,500$) spectra were obtained with a typical signal-to-noise
S/N$\sim 100$. Targets were spread over the main Galactic components,
including the halo \citep{zasowski2013,zasowski2017}. DR14 offers
elemental-abundance determinations for $\sim$19 chemical especies
determined by the APOGEE Stellar Parameters and Abundances pipeline
\citep[ASPCAP; ][]{garcia2016}.

We reproduce the analysis performed by \citet{fernandezalvar2017}, now
using the more extended sample of APOGEE-2 stars included in DR14. As in their
work, we select stars at distances from the Galactic plane, $|h|$,
larger then 5 kpc to avoid contamination from thick-disc stars. We consider
distances derived by the Brazilian Participation Group
\citep[BPG][]{santiago2016, queiroz2018} from the spectrophotometric
stellar atmospheric parameters (effective temperature, $T_{\rm eff}$, and
surface gravity, $\log g$) determined in DR14. 

We reject objects with
less reliable measured parameters, warned by the flags STAR\_BAD,
GRIDEDGE\_BAD, PERSIST\_LOW, PERSIST\_MED, PERSIST\_HIGH, TEFF\_BAD,
LOGG\_BAD,METALS\_BAD, ALPHAFE\_BAD, CHI2\_BAD, SN\_BAD, and
NO\_ASPCAP\_RESULT in the catalogue. In addition, we only consider stars
with spectra having S/N $> 80$, $T_{\rm eff} > 4000$ K, and $1.0 < \log
g < 3.5$.

These were the selection criteria chosen by \citet{fernandezalvar2017} for analysing the APOGEE DR12, to which we aim to compare our inferred chemical trends. Thus, we would like to maintain the same $T_{\rm eff}$ and $\log g$ selection cuts for consistence. Besides, although improvements in the ASPCAP were performed in DR14 (see \citet{holtzman18}), chemical issues have been still detected related to stars at lower $T_{\rm eff}$ and large $\log g$. For instance, \citet{zasowski19} revealed a strange high [O/Fe] trend at Solar metallicities in stars with $T_{\rm eff} < 3800$ K. And due to the lack of surface gravity estimations from asteroseismology at the time DR14 was released, calibrated $\log g$ values for stars at $\log g > 4$ are still not provided in DR14. 

We also exclude stars likely belonging to clusters. We do not exclude
targets in streams, because substructures will also be considered in the
simulation analysis. 

Our final sample comprises 1185 stars. The mean uncertainties estimated and their standard deviations, for the stellar parameters chemical abundances and distances in this stellar sample are the following: $\sigma(T_{\rm eff}) = 83 \pm 17$ K, $\sigma(\log g) = 0.09 \pm 0.02$, $\sigma($[O/Fe]$) = 0.07 \pm 0.04$, $\sigma($[Mg/Fe]$) = 0.05 \pm 0.02$, and $\sigma(d_{\rm sun})/d_{\rm sun} = 12 \pm 3 \%$. 

\subsection{Simulations}
We analysed six Milky-Way mass-sized haloes from the Aquarius project
\citep{scan09}. The initial conditions of the Aquarius haloes are
consistent with a $\Lambda$CDM model with $\Omega_{m} = 0.25$,
$\Omega_{\Lambda} = 0.75$, $\sigma_{8}=0.9$, $n_{S}=1$, and $H_{0}=100
h$ km $\rm s^{-1}$ Mp$\rm c^{-1}$ with $h = 0.73$. Dark matter particles
have masses on the order of $\sim 10^{6} M_{\odot}h^{-1}$; initially the
particles have masses of $\sim2 \times 10^{5} M_{\odot} h^{-1}$. The
analysed haloes correspond to level-5 resolution. The analysed
  haloes are Aq-C-5 and Aq-D-5. Hereafter, we will use them Aq-C and
  Aq-D for simplicity. 

The simulations were performed using a modified version of GADGET-3
\citep{scan05,scan06} that includes the physics of the baryons, chemical
evolution, and galactic mass-loaded outflows triggered by SNe explosions. The
chemical model follows the enrichment from both SNeII and SNeIa. 

The galactic systems are identified using the Friends-of-Friends
algorithm, and have virial masses in the range $\sim 5-11 \times 10^{11}
M_{\odot}h^{-1}$. Within the virial radius the simulations are
numerically resolved using $\sim 1$ million total particles at z=0.
Satellite galaxies that could be identified by the SUBFIND
\citep{springel2005} were taken out. Streams and substructures were not
removed. 

To define the stellar haloes, we applied the same criteria described in
\citet{tissera2012}, with an extra condition on the minimum height ($|h|$)
above the disc plane. We define the disc components by considering
particles dominated by rotation (i.e., using the ratio ($\epsilon$)
between the angular momentum along the z-axis and the total angular
momentum at that binding energy, so that star particles with
$\epsilon>0.65$ are taken as part of the disc components). Star particles
that are dominated by dispersion and have a height $|h| < 5$ kpc are
excluded from the analysis. This condition allows us to better mimic
the observations that, as described above, apply a similar condition
\citep[see also][]{carollo2018, monachesi2018}. For this analysis we 
consider the entire stellar haloes (i.e., no separation between inner and
outer haloes is applied).

In order to explore in more detail the origin of the $\alpha$-element
patterns in the inner region (within 30 kpc) of the stellar haloes, we adopt the
same classification as \citet{tissera2013}: (1) Stars born from gas accreted
in the first stages of assembly, (2) Disc-heated stars formed in the disc
structure of the main progenitor galaxy, then heated kinematically, and
(3) Stars formed from gas carried in by gas-rich satellite galaxies
('endo-debris').


\begin{figure*}
    \includegraphics[scale=0.8]{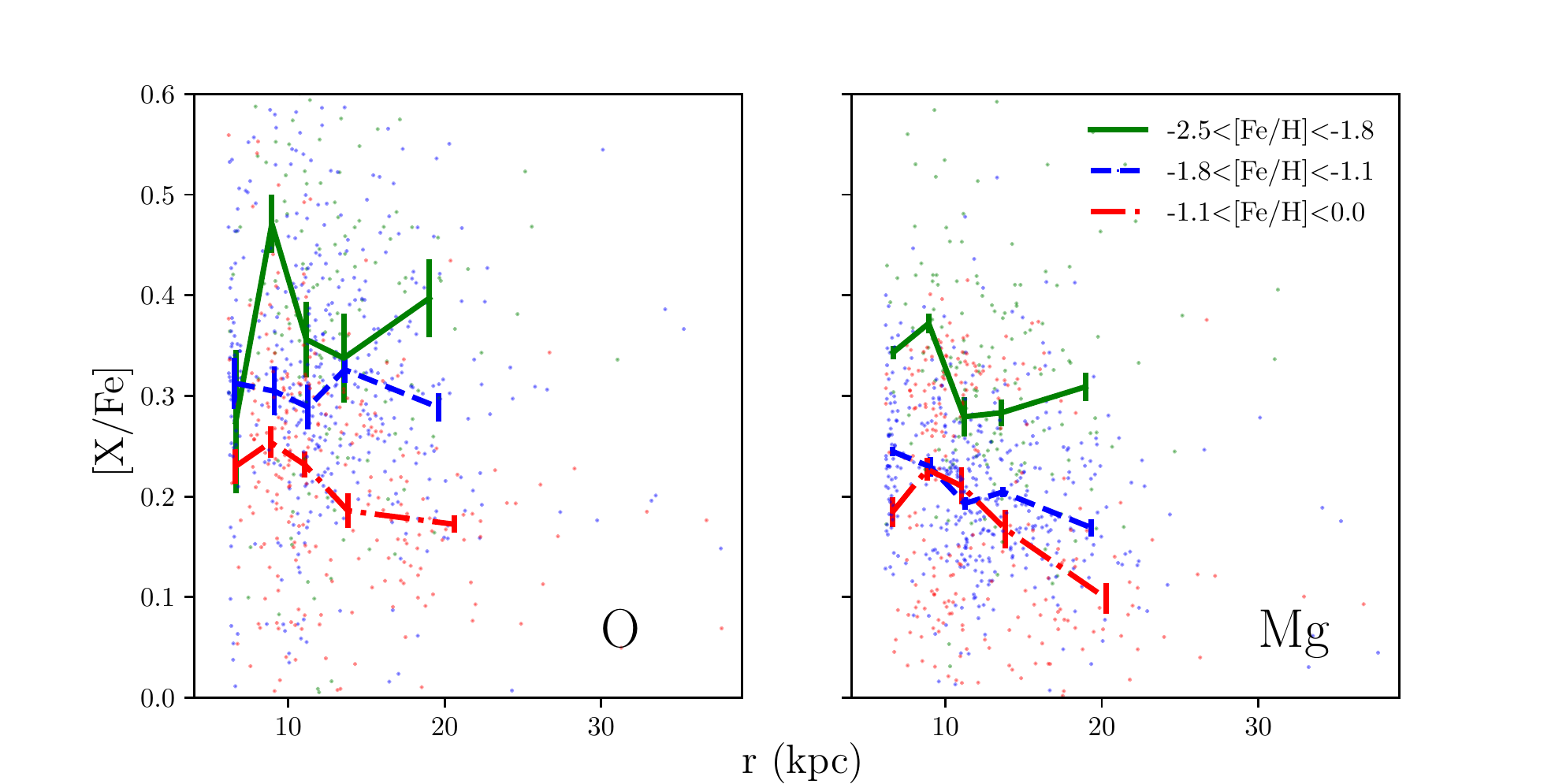}
    \includegraphics[scale=0.8]{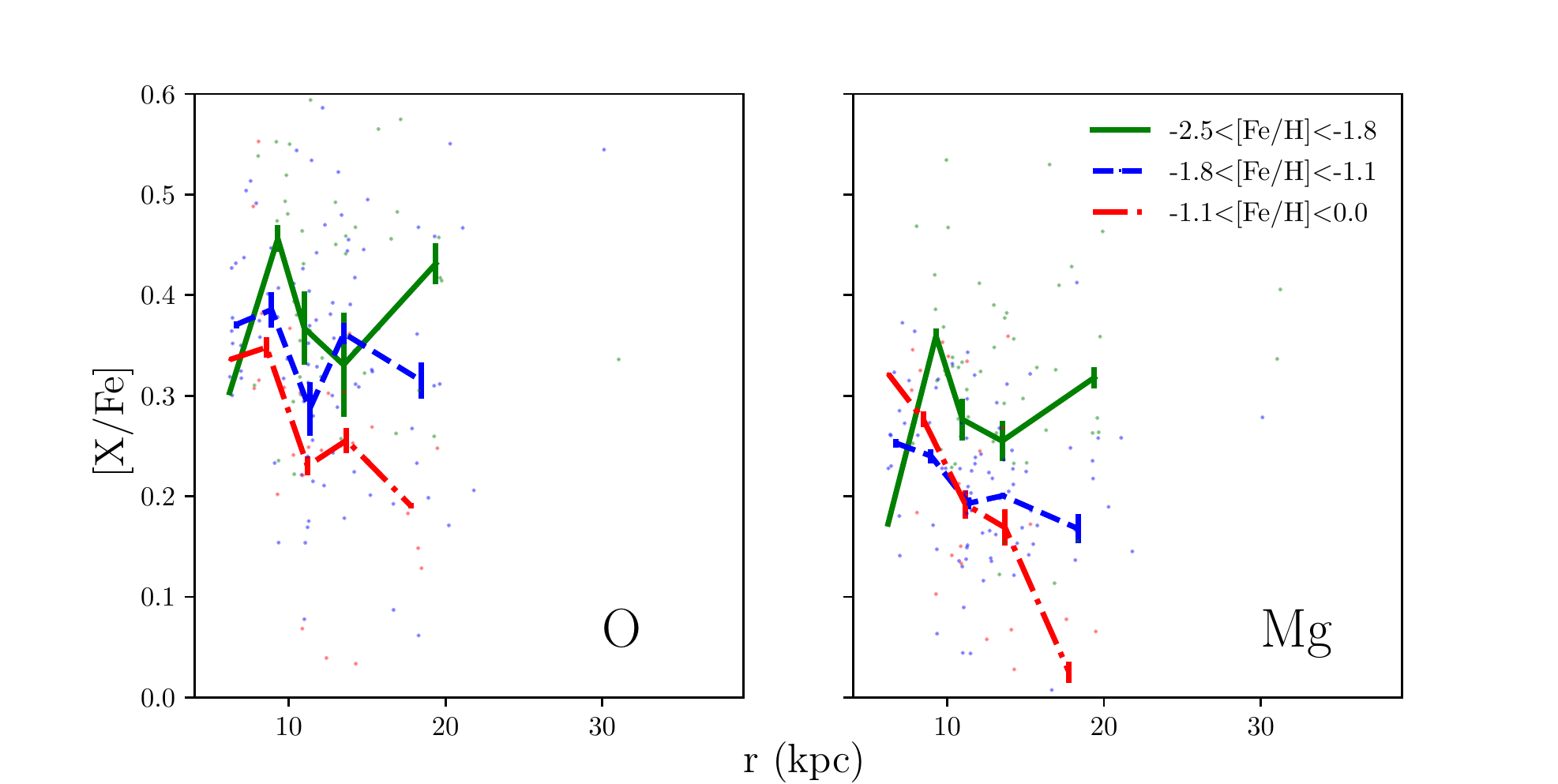}
        \caption{Top panels: Median [O/Fe] and [Mg/Fe] ratios determined for the selected whole sample in bins $r$ (from 5 kpc to
    30 kpc) for stars in three different ranges in [Fe/H]
    ($-2.5 <$ [Fe/H] $< -1.8$, $-1.8 <$ [Fe/H] $< -1.1$, and -$1.1 <$
    [Fe/H] $< 0.0$). The error bars are the statistical error for each median
    determined, the median absolute deviation (MAD). Bottom panels: Same as top panels but considering only stars with 4600 < $T_{\rm eff}$ < 4800 (reduced sample).}
    \label{figure1}
\end{figure*}

\begin{figure*}
	\includegraphics[scale=0.8]{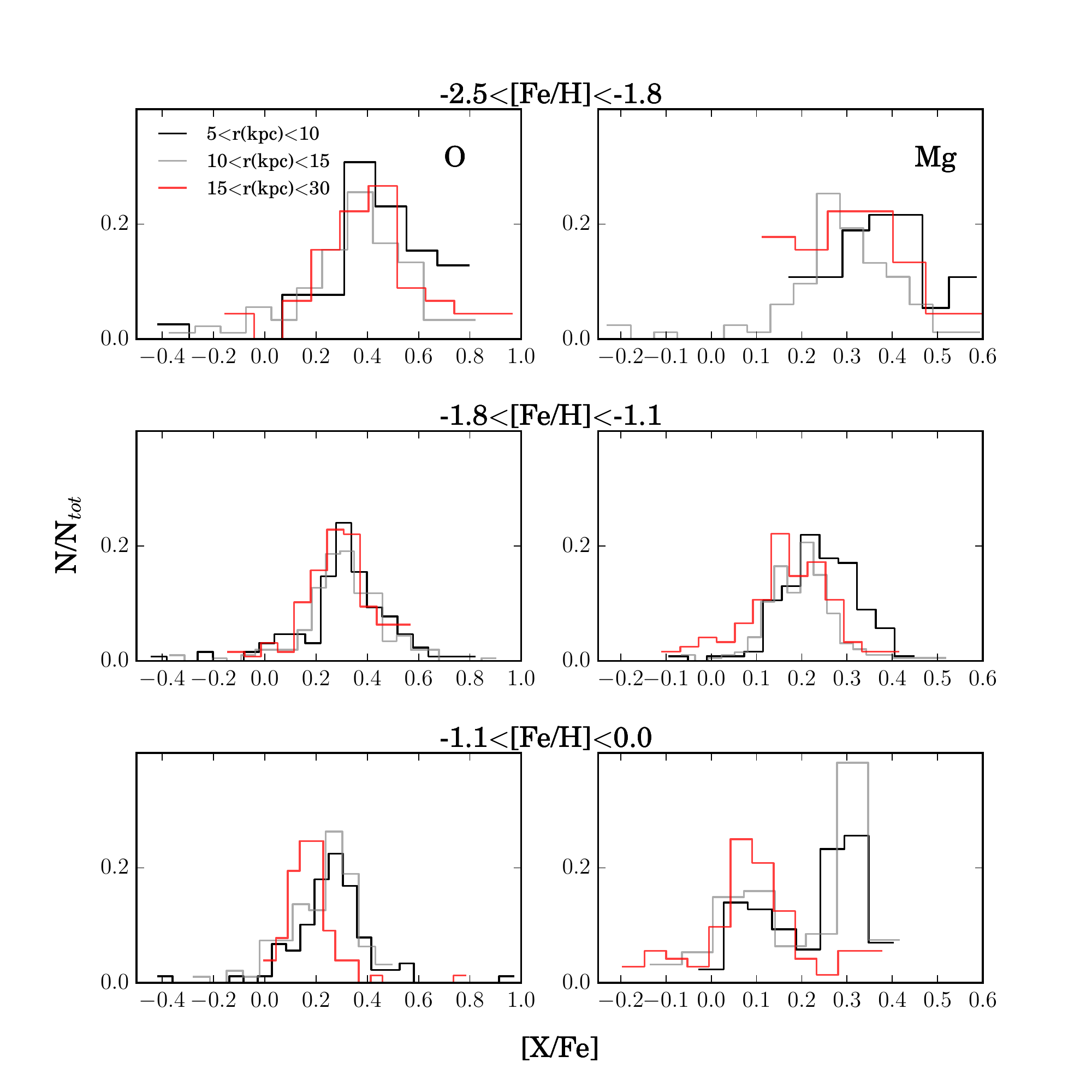}
	\caption{Distribution of [X/Fe] of the whole sample for each $r$ range (each histogram: 5 $< r <
    10$ kpc in black, 10 $< r <$ 15 kpc in grey, and 15 $< r < 30$ kpc in red)
    at the three different [Fe/H] bins considered, from the first to
    the third rows of panels, respectively. The y-axis corresponds to the number of stars ($N$) in each [X/Fe] bin, normalized to the total number of stars ($N_{tot}$) in each bin of $r$ and [Fe/H].}
    \label{histo_norm}
\end{figure*}

We acknowledge the fact that these simulated haloes appear to be more
massive than suggested by observations \citep{harmsen2017}. In part,
this could be due to an excess of in-situ stars or a misclassification
of thick-disc stars as disc-heated stars. Depending on the halo, the 
in-situ fraction contributes about 30 per cent of the inner stellar
halo \citep{tissera2012}. A better agreement between observations and
simulations is found if the in-situ components are not considered
\citep{carollo2018}. This is also the case for other simulations that
have been performed with different numerical codes and baryonic physics
\citep[e.g.][]{monachesi2018}. Nevertheless, the comparative analysis of the assembly
histories and the $\alpha$-element patterns of these two haloes provides useful clues 
for the interpretation of the current observations, as they have been performed with the same subgrid physics and numerical resolution, being the initial conditions (i.e. assembly histories) the only difference between them. 

In this
paper, we focus on the simulated [O/Fe] distributions as a function of
radius, which we take to be representative of [Mg/Fe] as well. The yields adopted for SNII are those of \citet{woosleyweaver1995} and they have some issues with Mg so that they produce less elements than expected. However the O abundances are well-reproduced.
As a consequence we decided to use O abundances for the simulations.

\section{The analysis}

\subsection{Observed chemical patterns}

We aim to properly study the [X/Fe] abundance variations across the halo
for $\alpha$-elements, as a function of metallicity, which we
identify with the [Fe/H] abundance ratio. We perform this analysis for
the [O/Fe] and [Mg/Fe] abundances released in DR14. These two elements
have a similar origin, and are mainly released into the interstellar medium
through the explosion of Type II supernovae. 
	
As in \citet{fernandezalvar2017}, we calculate the median [X/Fe] from
the abundances available in the DR14 catalogue in bins of distance from
the Galactic center, $r$, of 2.5 kpc, assuring a minimum of 5 stars per
bin. We calculate the medians for different metallicity ranges: $-2.5 <$
[Fe/H] $< -1.8$ (low-metallicity population), $-1.8 <$ [Fe/H] $< -1.1$
(intermediate-metallicity population), and $-1.1 <$ [Fe/H] $< 0.0$
(high-metallicity population).

The derived chemical trends for our entire halo sample are shown in
Figure \ref{figure1}. The panels exhibit the resulting [X/Fe] medians,
as well as their corresponding median absolute deviation (MAD), as a function
of $r$. 

The chemical trends for both elemental-abundance ratios exhibit a
significant decrease with $r$ for stars at $-1.1 <$ [Fe/H] $< 0.0$. 
We first evaluate the possibility that this gradient is the result 
of our selection criteria. We obtain the $\log g$ distribution
at each [Fe/H] and radial bin. We note that for $r > 15$ kpc the $\log
g$ distribution is skewed toward lower values (peak at $\log g \sim
1.0$), whereas at lower distances the distribution peaks at higher
$\log g$ values. The [O/Fe] abundances do not show a trend with $\log
g$, except a slightly increasing trend with $\log g$ in the $-1.1 <$
[Fe/H] $<0.0$ bin at $r > 15$ kpc. Stars in this bin exhibit lower [O/Fe]
values at $\log g \sim 1.0$ than stars at $r < 15$ kpc. However, this
trend with $\log g$ is not observed in the case of [Mg/Fe], for which
the decreasing trend of this ratio with distance is the same as observed 
for [O/Fe]. For this reason, we consider that the decrease detected
for the [$\alpha$-elements/Fe] ratio with distance from the Galactic
center is not due to the shift in the $\log g$ distribution toward lower
values at the largest distances.

The oxygen abundances are determined from 50 regions of the spectra
covering mostly OH molecular bands (Holtzman et al. 2018). The OH bands are very sensitive to the adopted $T_{\rm eff}$. We
detect a bias in [Fe/H] towards high values at $T_{\rm eff} >$ 4600\,K.
We also detect an increasing trend of [O/Fe] with $T_{\rm eff}$, and an
important increase in the abundance dispersion for stars with $T_{\rm eff}
> 5000$\,K. [Mg/Fe] abundances also show an increasing trend with
$T_{\rm eff}$, although much lower than for [O/Fe]. In order to avoid
artificial distortions in the derived trends with distance due to these
systematic effects, we reduce our sample to a narrower range of $T_{\rm
eff}$, [4600 , 4800]\,K, where there is no bias toward higher [Fe/H] and no dependence of the chemical abundances with $T_{\rm eff}$ is observed. This reduced sample comprised 183 stars. The bottom panels in Fig. \ref{figure1} show the resulting trends inferred from this smaller group. Due to the lower number of stars, the dispersion in the resulting median trends is larger and the statistical results are less robust. However, the trends inferred are
qualitatively the same as those derived from the whole sample, in particular the steeper decreasing trend observed in the most metal-rich bin. The fact that we detected the same patterns with $r$ using a sample where the chemical abundances do not show systematic correlations with stellar parameters suggests that these trends are real, and not an artifact.

Figure \ref{histo_norm} shows the [X/Fe] distribution in each of the
[Fe/H] ranges considered, and three radial bins, $5 < r < 10$ kpc,
$10 < r < 15$ kpc,  and $15 < r < 30$ kpc, normalised to the
total number of stars in each radial bin. 
	
These trends, inferred from a larger number of halo stars and improved
abundance determinations and distances, confirm the decrease in
[$\alpha$/Fe] with $r$ for stars at [Fe/H] $> -1.1$ reported by
\citet{fernandezalvar2017}.

\subsection{Simulated chemical trends}

\begin{figure}
    \includegraphics[scale=0.4]{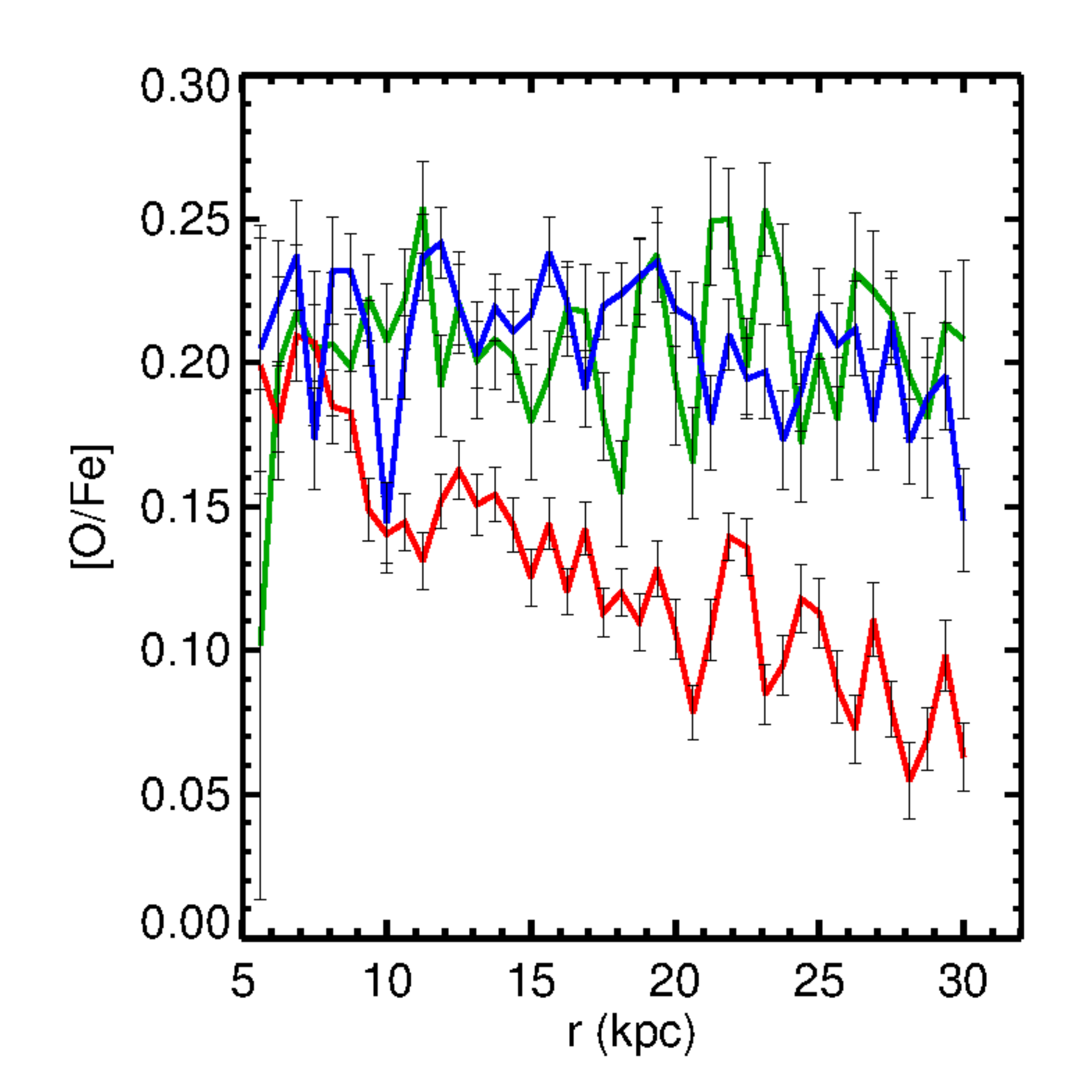}
    \includegraphics[scale=0.4]{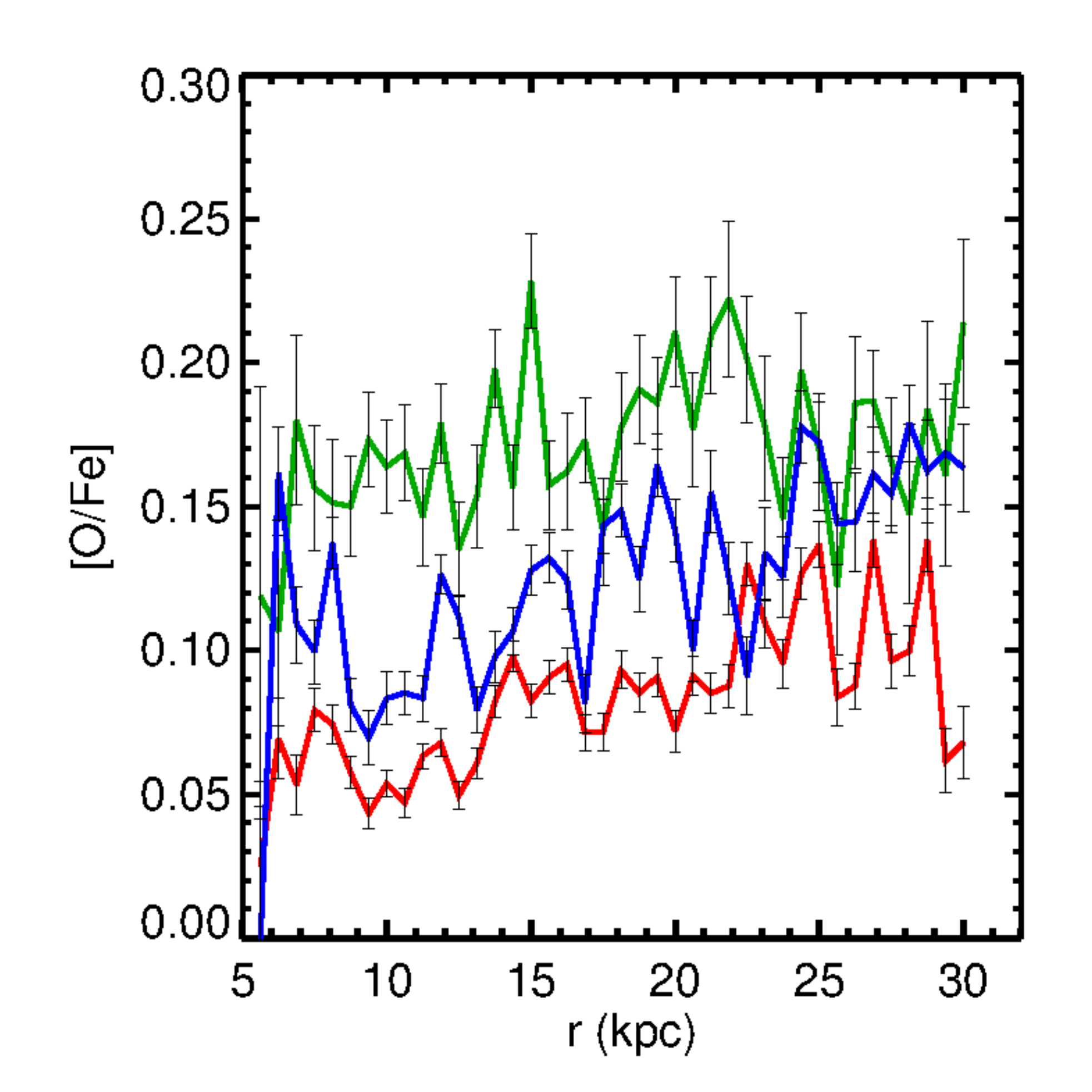}
        \caption{Median [O/Fe] values derived from the Aq-C (top panel)
    and the Aq-D (bottom panel) halos obtained from the Aquarius
    cosmological simulations, in the same [Fe/H] and $r$ bins, and using
    the same color coding, as the observations.} \label{figure2}
\end{figure}

We estimated the [O/Fe] profiles for the seven Aquarius haloes (A, B, C,
D, F, G, H) analysed by \citet{tissera2012}, calculating medians in the
same [Fe/H] and $r$ intervals used for the observations. We
only considered stars located at $|h| > 5$ kpc, the same requirement
imposed for the APOGEE observations. In the case of the simulations, 
$^{16}$O is our $\alpha$-element reference. After exploring the
trends in all the haloes, we choose two of them to make a detailed
analysis in relation to their assembly histories, Aq-C and Aq-D. 
The simulated halo that best represents the observational
[O/Fe]  and [Mg/Fe] trends shown in Fig. ~\ref{figure1} is Aq-C, which exhibits a
decrease in these trends with $r$ for the most metal-rich bin, stars
with [Fe/H]$ > -1.1$. On the contrary, the halo that differs the most from
the observations is Aq-D, exhibiting a slightly increasing
$\alpha$-enrichment with increasing radius. We show the trends for both
haloes in Fig.\ref{figure2}. From these figures, we also note that, for
the intermediate-[Fe/H] interval, Aq-D also exhibits an increase of
[O/Fe], while Aq-C has a flat trend with radius. In the low-[Fe/H] interval, the
trends with radius are flat for both simulated haloes.

\begin{figure*}
\includegraphics[scale=0.8]{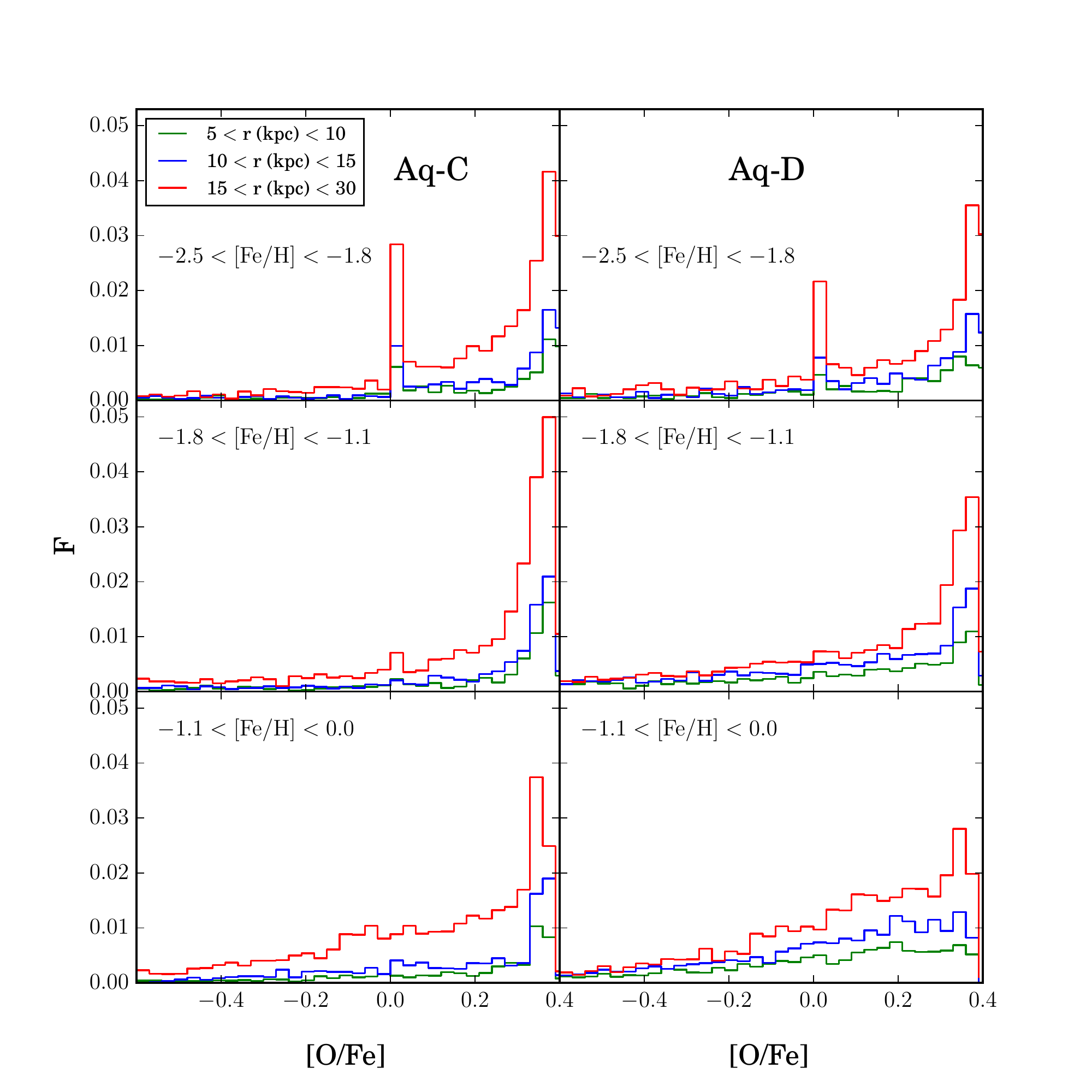}
\caption{Mass fractions of stars within the three metallicity intervals:
$-2.5 <$ [Fe/H] $< -1.8$ (upper), $-1.8 <$ [Fe/H] $< -1.1$ (middle), and
$-1.1 <$ [Fe/H] $< 0.0$ (lower panels) and in three radial intervals: $5
< r < 10$ kpc (blue lines), $10 < r < 15$ kpc (green lines), and $15 < r
< 30$ kpc (red lines), for haloes Aq-C and Aq-D (left and right columns,
respectively), normalized to the total mass in each bin of $r$ and [Fe/H]. }
\label{histosimus}
\end{figure*}

In order to further compare the observations and simulations, 
Fig.~\ref{histosimus} shows the [O/Fe] distributions for the defined
radial intervals. At first sight, the distributions appear different from
the observed ones shown in Fig.~\ref{histo_norm}\footnote{The sharp
increase in [O/Fe] close to Solar values is produced by the enrichment from
SNIa relative to SNII. This occurrs in the low-metallicity sub-samples were
the star formation is starting, and the mixing of chemical elements in
the ISM is more inhomogeneous. More efficient metal mixing will solve
this numerical artefact.}. However, when examined in
detail, some similarities emerge. In the case of Aq-C, the trends from the
the inner to outer radii for the three metallicity sub-samples are similar:
the inner radii are more  dominated by $\alpha$-rich stars with
respect to the 
$\alpha$-poor stars, principally   in the outer radial interval.
Althougth at all radii and metallicities, there is a larger
contribution of $\alpha$-rich stars in Aq-C, the relative contribution of
$\alpha$-poor stars increases remarkably for high metallicities in
outer radial interval. It is this different relative contribution of
$\alpha$-poor stas that results in a slope more comparable to the
MW abundance patterns.
A similar change in the relative contribution of
$\alpha$-rich to $\alpha$-poor stars can be seen in Fig. \ref{histo_norm}. However, we
acknowledge the fact that the observations show a sharper decrease of
$\alpha$-rich and high-metallicity stars in the outer region than that
predicted by Aq-C.

Conversely, Aq-D does not exhibit such a significant contribution from
$\alpha$-rich stars in the high-metallicity interval at any radius,
because there is a larger contribution of $\alpha$-poor stars at all
radii. In addition, the contribution of $\alpha$-rich stars increases
for $r > 15$ kpc. This is why the median [O/Fe] in the central regions
of Aq-D are lower than in the outer radial interval, producing the
opposite trend to that found in Aq-C. Below we explore why this happens
in Aq-D but not in Aq-C, the simulated halo that is most like the
observed MW [O/Fe] pattern.

\begin{figure*}
\includegraphics[scale=0.8]{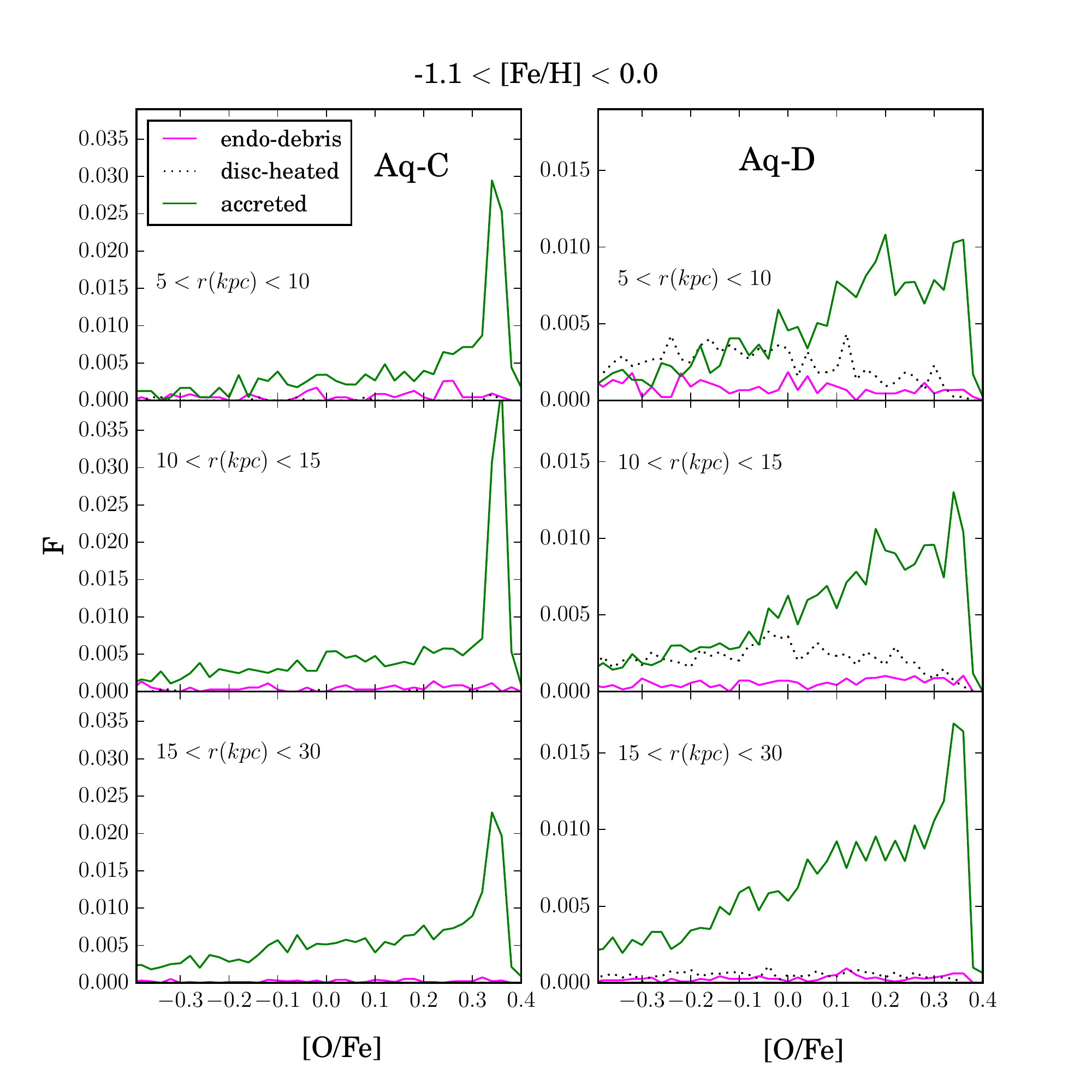}
\caption{Mass fractions of stars, F, at $-1.1 < $ [Fe/H]$ < 0.0$ in the three radial intervals: 
$5 < r < 10$ kpc (green lines), $10 < r < 15$ kpc (blue lines), and $15 < r
< 30$ kpc (red lines), normalized to the total mass in this [Fe/H] bin, for haloes Aq-C and Aq-D (left and right columns,
respectively), considering separately each sub-population with a
different origin. }
\label{figure3}
\end{figure*}

\subsection{The high-metallicity stellar population}
To understand which stellar populations (according to their origin) are
responsible of the [O/Fe] gradient in stars with $-1.1 < $ [Fe/H] $ <
0.0$, we examine the contribution of each sub-population according to
their origin. We focus on this metallicity interval because it is the
one which shows the strongest signature in the observations (see
Fig.~\ref{figure1}).

Figure \ref{figure3} shows the contribution of the stellar populations
to the [O/Fe] distributions for those stars with $-1.1 < $ [Fe/H]$ < 0.0$,
divided into the three defined radial intervals for the Aq-C and Aq-D
simulated haloes. Each sub-sample has been normalised to the total
stellar mass within a radial interval. This figure reveals that the Aq-C
halo is characterised by a conspicuous fraction of accreted stars with
high-[O/Fe] ($\sim +0.3$) for each $r$ interval. The contribution of
in-situ stars is almost negligible compared with the contribution of
accreted stars. The fraction of accreted stars with [O/Fe] $< +0.2$
increases with $r$, whereas that corresponding to high [O/Fe]
decreases, causing the negative [O/Fe] gradient with $r$. Therefore, the
stellar population responsible for the negative [O/Fe] gradient with
distance is the large fraction of accreted stars with low
$\alpha$-values.

In Aq-D, there is smaller relative contribution of accreted high-[O/Fe] 
stars. There is also a larger contribution of disc-heated stars with
low [O/Fe]. This fraction of in-situ stars decreases as $r$ increases,
so the contribution of stars with high [O/Fe] increases with $r$.
In addition, the distribution of [O/Fe] is broader, without a strong peak at
high [O/Fe], as in the case of Aq-C. The relative fraction of high-[O/Fe] stars
tend to increase with $r$. Consequently, the median [O/Fe] value for
stars within this metallicity range is higher, as they are located farther
away. In this simulated halo, the $\alpha$-poor stars are more
concentrated in the inner regions.


\subsection{The assembly history behind the $\alpha$-element trends}
In the case of Aq-C, the distribution of [O/Fe], with a peak at high
[O/Fe] $\sim +0.35$ at high metallicities ([Fe/H] $> -1.1$), implies that
these stars should have formed from a short and intense burst of star
formation. The ISM would have reached large [Fe/H] and [O/H], both
mainly due to the contribution of SNeII. The stars contributing to the
nearer regions ($r < 15$ kpc) should be old and contemporaneous,
otherwise SNeIa would have had time to explode, enriching the ISM so
that the subsequently born stars would have lower [O/Fe]. On the other
hand, the increase of low [O/Fe] at more distant radii implies that these
stars would be comparatively younger.

The broader distribution of [O/Fe] displayed by the accreted sample in
the halo Aq-D, with a more similar fraction of low-[O/Fe] stars with respect
to high-[O/Fe] stars, indicates that they would have formed during an
extended star-formation era, allowing for a significant fraction of
stars to be formed from gas enriched by SNeIa.

\begin{figure*}
\includegraphics[scale=0.8]{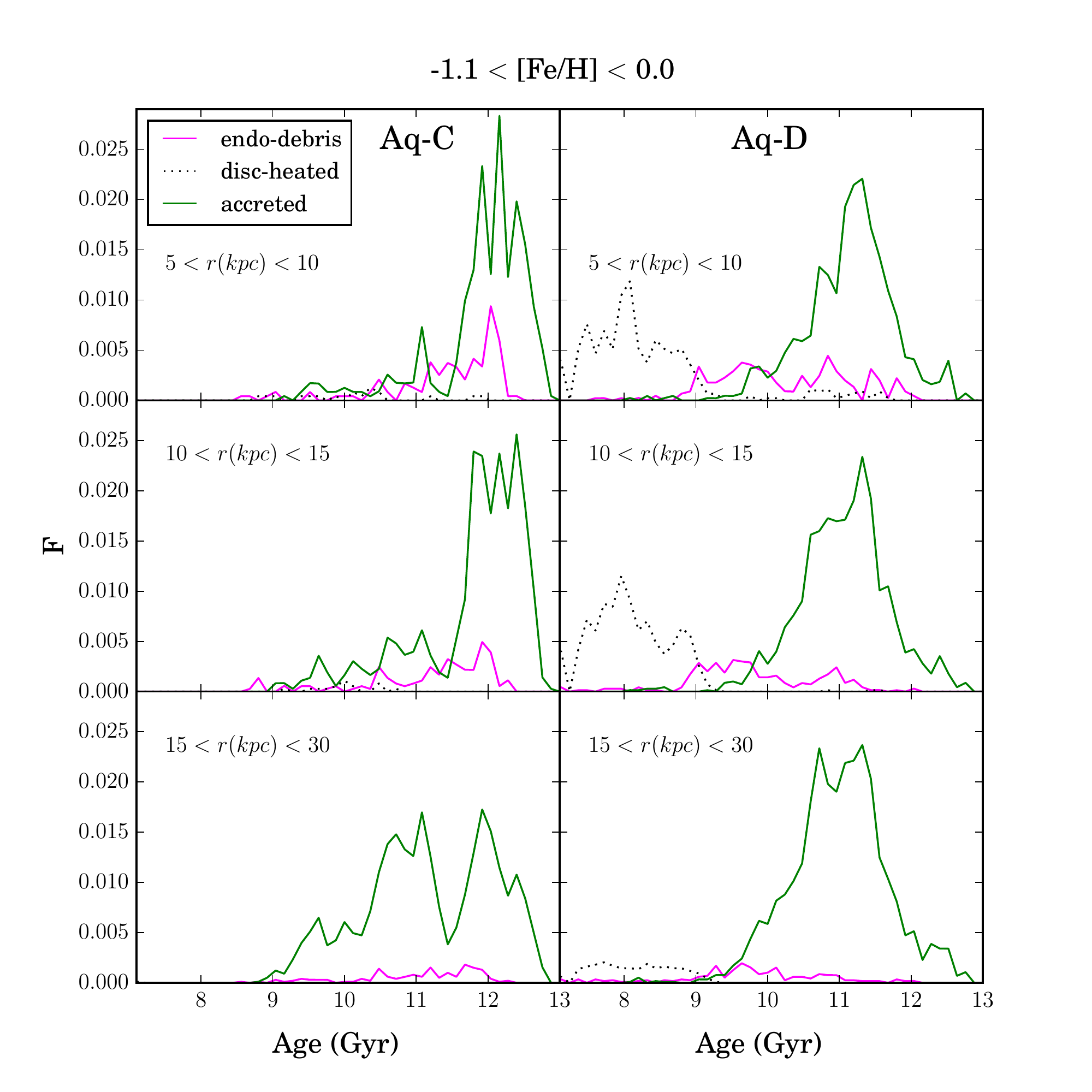}
\caption{Distribution of simulated ages from the Aq-C and Aq-D haloes
for each stellar population with different origin and  [Fe/H] > -1.1. 
We show the distributions for each adopted radial interval as in Fig.~\ref{figure3}.}
\label{figure4}
\end{figure*}

To further probe this hypothesis, we examine the age distribution of each
stellar population. As can be seen from Fig. \ref{figure4}, the accreted
stars in the halo Aq-C populating the nearer region are concentrated in
an age range [$\sim11.5, 13.0$] Gyr. A secondary peak at $\sim 11$ Gyr,
which would correspond to stars with lower [O/Fe] values, is
considerably lower with respect to the rest of the distribution, but
increases with increasing radii and extends towards lower ages. This age
distribution explains the increment of low-[O/Fe] stars, which are
indeed slightly younger stars, formed at an epoch when the ISM would
have been already enriched by SNeIa. There is also a fraction of in-situ
endo-debris stars peaked at $\sim 12$ Gyr for stars at $r < 10$ kpc,
which diminishes at higher radii. 

Conversely, the distribution of stellar ages for the halo Aq-D peaks at
slightly lower values, $\sim 11.5$ Gyr, and extends over the range
[$\sim 9.5$, $\sim 13$] Gyr. These stars formed during a more extending
period compared with Aq-C. From Fig.~\ref{figure5} we can also see that
there are disc-heated stars with ages in the range [$\sim 7$, $\sim
9.5$] Gyr. In the intermediate radial range, there is also an
significant contribution from disc-heated stars in this halo.

The main difference between the age distributions of the
high-metallicity stars in Aq-C and Aq-D is that the former has older
stellar populations in the inner regions, and in the outer radial
interval the age distributions are consistent with two star-formation
episodes. This is not the case for Aq-D, which exhibits a more extended
period of star formation in all three radial intervals.

\subsection{Accreted satellites}
\begin{figure*}
\includegraphics[scale=0.8]{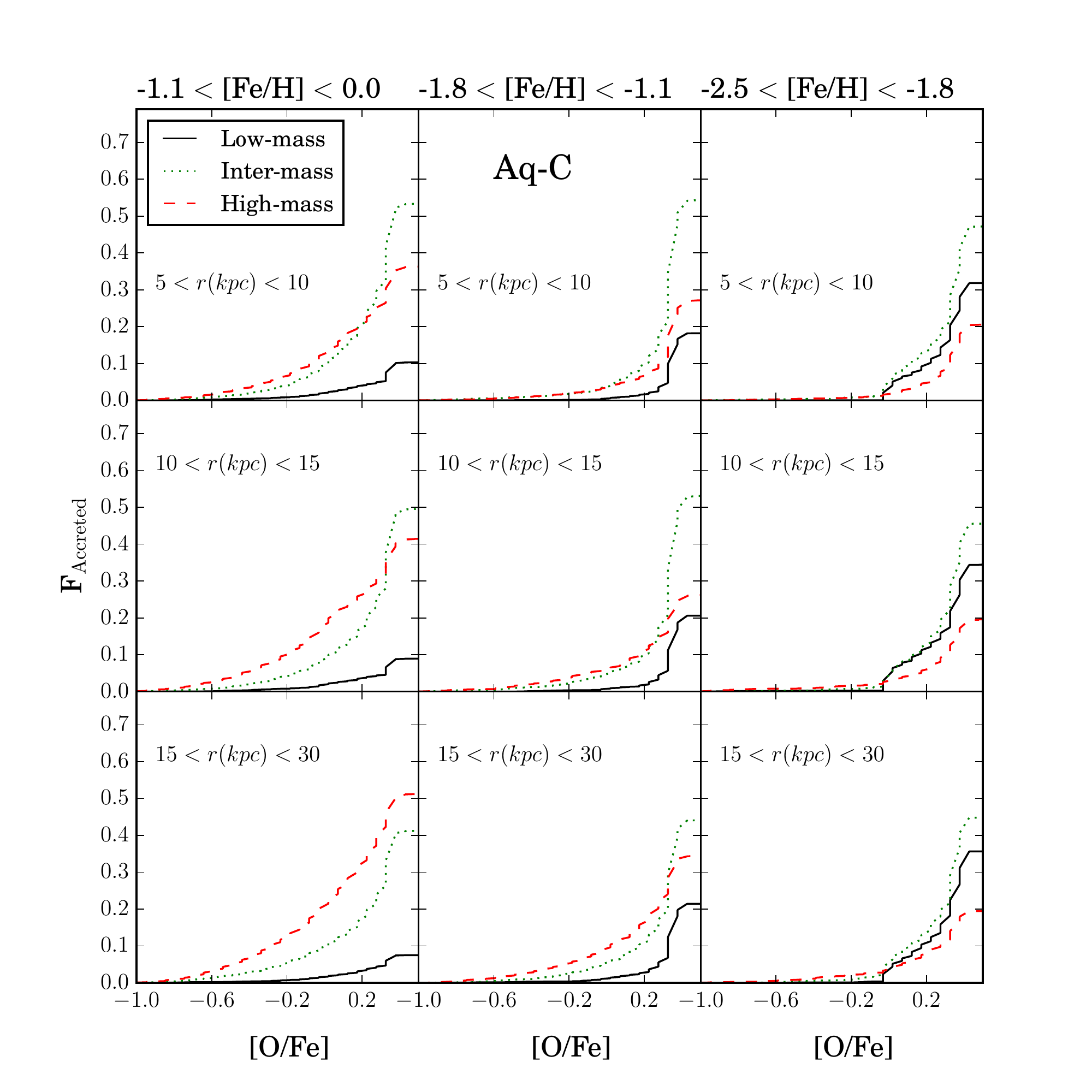}
\caption{Mass-weighted distributions of the accreted stellar
  populations of Aq-C, grouped according to the mass of their host
satellites, normalized to each $r$ and [Fe/H] bin. Accreted stars are shown according to different metallicity
intervals (left:high-metallicity, middle:
  intermediate-metallicity, right: low-metallicity), and different
radial intervals (upper: inner regions, middle: intermediate regions, lower panels:
  outer regions). }
\label{figure5}
\end{figure*}

\begin{figure*}
\includegraphics[scale=0.8]{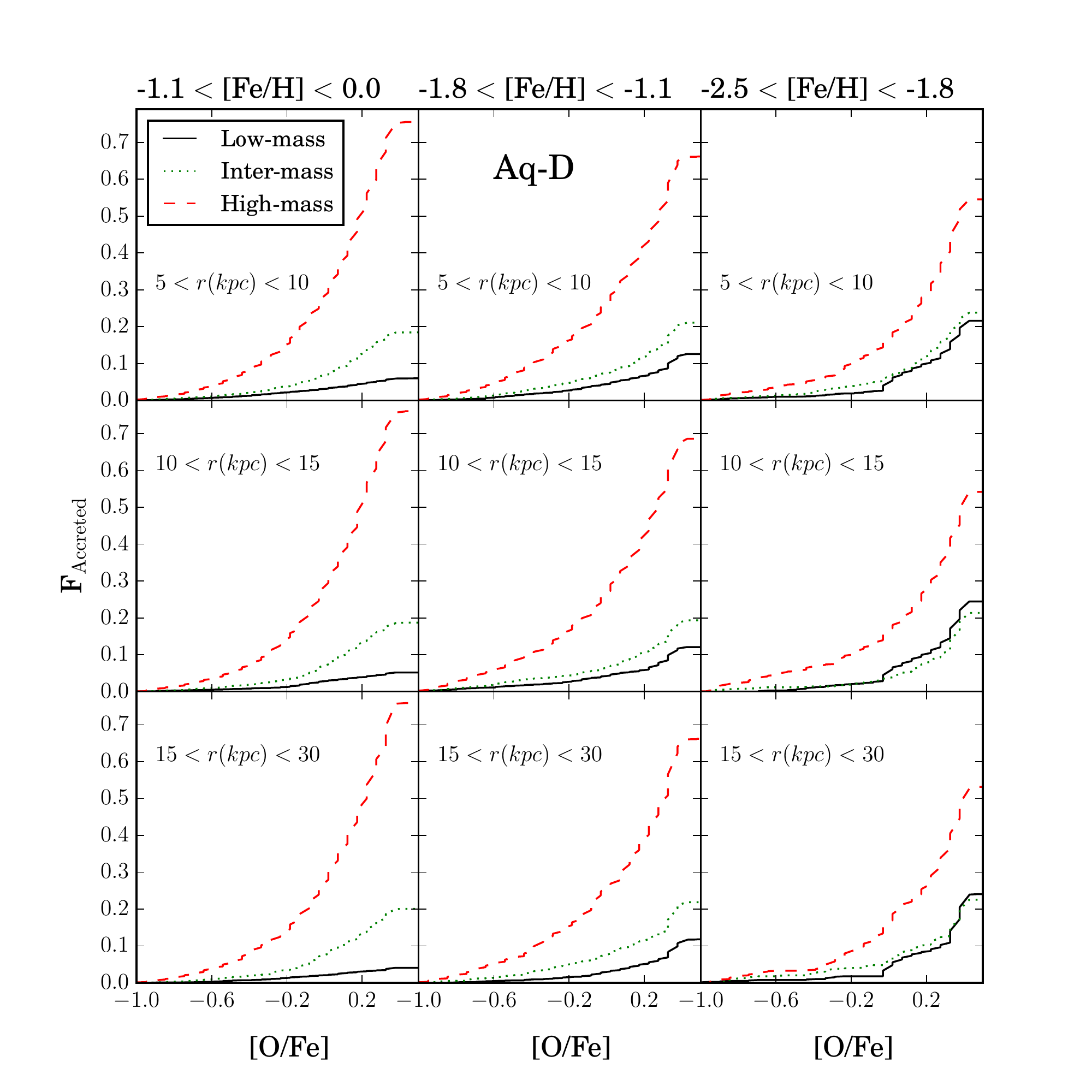}
\caption{Mass-weighted distributions of the accreted stellar
  populations of Aq-D, grouped according to the mass of their host
satellites, normalized to each $r$ and [Fe/H] bin. Accreted stars are shown according to different metallicity
intervals (left:high-metallicity, middle:
  intermediate-metallicity, right: low-metallicity), and different
radial intervals (upper: inner regions, middle: intermediate regions, lower panels:
  outer regions).}
\label{figure6}
\end{figure*}

To further understand the contributions of the accreted satellite
galaxies to the formation of the inner regions of these haloes and,
therefore, their role in setting the [O/Fe] profiles,
Fig.~\ref{figure5} and Fig.~\ref{figure6} show the contribution of
stars with different metallicities to the different radial intervals of
the inner stellar haloes, by small (M $< 10^{9}$M$_{\odot}$) ,
intermediate ($10^{9}$M$_{\odot} <$ M $<10^{10}$M$_{\odot}$), and massive
(M $>10^{10}$M$_{\odot}$) satellites (where M denotes the dynamical mass
of the accreted satellites at the time it enters the virial radius of
the main progenitor galaxy). 


As shown in Fig.~\ref{figure5}, Aq-C has a larger contribution from
intermediate-mass satellites for the lower metallicity interval. The
higher-metallicity accreted stars are formed in massive and
intermediate-mass satellites. For increasing radius, the contributions from massive
satellites increases. This is consistent with the previous trends
showing these stars have lower [O/Fe] abundances. From figure 4 in
\citet{tissera2014}, we know that the Aq-C inner region did not receive
satellites larger than about $\sim 2\times10^{10}$M$_{\odot}$. Our
analysis shows that three satellites, with dynamical masses not larger
than $2\times 10^{10}$ M$_{\odot}$, contribute stars particularly
in the outer radial interval.

Regarding Aq-D, the trends in Fig.~\ref{figure6} show a different
assembly history, with a more significant contribution coming from massive
satellites, also in agreement with figure 4 in \citet{tissera2014}.
These satellites are more massive, and contribute to all radial and
metallicity intervals. These more massive satellite accretion affects the
inner region and contributes less $\alpha$-enhanced stars to the
inner radial bins. We identify contributions from up to four satellites
with mean dynamical masses $\sim 2.5\times10^{10} $M$_{\odot}$ (i.e., in
the range [$\sim 2.5$ , $\sim 4.3\times10^{10}$] M$_{\odot}$).


For both Aq-C and Aq-D, these satellites did not deposit all of their stars
within these regions. There are partial contributions as the satellites
are disrupted. In the case of Aq-D, the more-massive satellites
contribute $\sim 75\%$, $\sim 66\%$, and $\sim 56\%$ of the stars in
the high-, intermediate-, and low-metallicity populations. Conversely, for
Aq-C, the higher-mass satellites contribute $\sim 39\%$, $\sim
28\%$, and $\sim 21 \%$ for the same metallicity populations. However,
this halo has the largest contributions coming from intermediate-mass
satellites:$\sim 52\%$, $\sim 53\%$, and $\sim 47\%$ of the stars in the
high-, intermediate-, and low-metallicity populations, respectively.


In summary, Aq-C shows a similar [$\alpha$/Fe] trend as a function of radius to the observations; it has been already reported that a late-infalling massive satellite with an extended SFH is contributing significant mass at intermediate distances -- for a very detailed description of that satellite orbit see \citet{parry2012} and Appendix A1 of \citet{cooper2017}.
Aq-D shows the opposite trend, supporting the idea that incomplete mixing of a massive progenitor is responsible for what is observed in the MW.

In fact, considering recent observational results in the MW regarding the existence of remnants of accreted satellites in the stellar halo \citep{belokurov2018,helmi2018}, we carried out the exercise of selecting high-metallicity star particles with large radial velocities ($|v_{r}| > 150$ km/s) and small tangential ones $|v_{t}| < 30$ km/s, contributed by  satellites of different masses in order to find out clues on their origins. We find that, in Aq-C,  50, 32 and 18 per cent of high-metallicity stars were contributed by massive, intermediate and small mass galaxies. While for Aq-D these percentages are 73, 19, and 7 per cent, respectively, in agreement with the trends shown in Fig.6 and Fig.7.
It is clear that in both cases the majority of the high-metallicity stars are associated to massive satellites. However, in the case of Aq-D, the massive satellite  clearly has the principal role. 
Nevertheless, this last halo do not have a distribution of $\alpha$-elements consistent with observations in the MW because most of the stars are  $\alpha$-poor  and they have been able to reach the center. So that the $\alpha$-profiles is inverted with respect to that of Aq-C.
It is very interesting to take into account the results of
\citet{cooper2017} that tracked the satellite in time and found it
has gone several pass-bye in the last 8 Gyrs. We cannot prevent the
obvious question: in the case of the MW, could be the satellite
responsible of imprinting this $\alpha$-pattern still around \citep[e.g.][]{belokurov2018,koposov2018}?

\section{Conclusions}

The elemental-abundance ratios for the two $\alpha$-elements O and Mg
reported in
the DR14 APOGEE/APOGEE-2 database for relatively metal-rich halo stars
([Fe/H] $> -1.1$) at Galactocentric radii between 5 and $\sim$30 kpc exhibit a gradient
with distance, which becomes flatter at lower metallicity. The simulated
halo that best reproduces this observational feature is Aq-C. This halo
exhibits a decreasing trend in [O/Fe] with $r$ for stars at [Fe/H] $> -1.1$
and flat trends at lower [Fe/H], for stars at $r > 5$ kpc,
qualitatively similar to the [O/Fe] trends inferred from the
APOGEE/APOGEE-2 DR14 observations. 

The results from the comparative analysis of Aq-C and Aq-D indicates
that the decreasing [O/Fe] with radius for more metal-rich stars is due to
the contribution of accreted stars from satellite galaxies with about
$10^{10}$ M$\sun$ dynamical mass that did not get all way to the
central region. The assembly history for Aq-C is characterised by
the larger contribution from intermediate-mass satellites in the inner
regions ($r < 15$ kpc) with old high-$\alpha$ stars, and the increase of
the contribution from more-massive satellites, populating it with younger
low-$\alpha$ stars, at $r > 15$ kpc.

The large fraction of old high-[O/Fe] stars at $r < 15$ kpc is due
to short and intense bursts of star formation, while the low-[O/Fe]
stellar populations are associated with a later second burst. There is a
clear gap between the two starbursts, which provides enough time for
SNIa enrichment. Note that these two $\alpha$-enriched populations might have
formed in different accreted satellites.

An intermediate/massive satellite would be able to retain gas to have a
second starburst, after the first one quenched star formation
by heating up the ISM (Aq-C). However, if it is sufficiently massive,
then the period of star formation would be more extended, and the
satellite would also be able to reach the inner regions
\citep{amorisco2017a,fattahi2018} as is the case in Aq-D. The resulting
[O/Fe] patterns will be different: more enriched material would be able
to reach the central region in the latter case, thus producing a positive
increases of [O/Fe] with radius. Note that these satellites could also
heat up the old disc, contributing with more disc-heated stars in the
halo. Our results are in agreement with those recently reported in the
literature pointing to the accretion of a relatively massive
satellite. In particular, our results agree with \citet{mackereth2018}
who analysed simulated galaxies from the EAGLE project and inferred a
merger history for the Milky Way of satellites with stellar masses in
the range $10^{8}$-$10^{9}$M$_{\odot}$.

The comparative analysis between Aq-C and Aq-D we have carried out
suggests that, in order to reproduce the observed $\alpha$-element patterns in the inner
region of the MW, in particular the decrease of [O/Fe] for increasing
radius, intermediate- and low-mass satellites would be expected to
contribute to the very inner regions but a more-massive satellite
(i.e., dynamical mass about $\sim 10^{10}$ M$_{\odot}$ at the time
  the satellite enters the virial radius of the MW progenitor)
is expected to contribute low-$\alpha$ stars farther away. Considering our results and those of Cooper et al . (2017) we speculate that the surviving remnant of this massive satellite could be still orbiting the MW.

\section*{Acknowledgements}
We thank Andrew Cooper for his detailed comments.
E.F.A. acknowleges financial support from the ANR 14-CE33-014-01, and partial support provided by CONACyT of M\'exico (grant
247132). P.B.T. acknowledges financial support from UNAB project
D667/2015 and Fondecyt Regular 1150334. Part of the analysis was done in RAGNAR cluster of the Numerical Astrophysics group at UNAB. L.C. thanks for the financial
supports provided by CONACyT of M \'exico (grant 241732), by PAPIIT of M
\'exico (IG100115, IA101215, IA101517) and by MINECO of Spain
(AYA2015-65205-P). This project has received funding from the European
Union Horizon 2020 Research and Innovation Programme under the Marie
Sklodowska-Curie grant agreement No 734374. T.C.B. acknowledges partial
support from grant PHY 14-30152 (Physics Frontier Center/JINA-CEE)
awarded by the U.S. National Science Foundation (NSF).

Funding for the Sloan Digital Sky Survey IV has been provided by the
Alfred P. Sloan Foundation, the U.S. Department of Energy Office of
Science, and the Participating Institutions. SDSS-IV acknowledges
support and resources from the Center for High-Performance Computing at
the University of Utah. The SDSS web site is www.sdss.org.

SDSS-IV is managed by the Astrophysical Research Consortium for the
Participating Institutions of the SDSS Collaboration including the
Brazilian Participation Group, the Carnegie Institution for Science,
Carnegie Mellon University, the Chilean Participation Group, the French
Participation Group, Harvard-Smithsonian Center for Astrophysics,
Instituto de Astrof\'isica de Canarias, The Johns Hopkins University,
Kavli Institute for the Physics and Mathematics of the Universe (IPMU) /
University of Tokyo, Lawrence Berkeley National Laboratory, Leibniz
Institut f\"ur Astrophysik Potsdam (AIP), Max-Planck-Institut f\"ur
Astronomie (MPIA Heidelberg), Max-Planck-Institut f\"ur Astrophysik (MPA
Garching), Max-Planck-Institut f\"ur Extraterrestrische Physik (MPE),
National Astronomical Observatories of China, New Mexico State
University, New York University, University of Notre Dame,
Observat\'ario Nacional / MCTI, The Ohio State University, Pennsylvania
State University, Shanghai Astronomical Observatory, United Kingdom
Participation Group, Universidad Nacional Aut\'onoma de M\'exico,
University of Arizona, University of Colorado Boulder, University of
Oxford, University of Portsmouth, University of Utah, University of
Virginia, University of Washington, University of Wisconsin, Vanderbilt
University, and Yale University.




\bibliographystyle{mnras}

\bibliography{bibliography} 

\bsp	
\label{lastpage}
\end{document}